\documentclass[aps,prb,twocolumn,showpacs,floats,preprintnumbers]{revtex4}
\usepackage{epsfig,amsopn}
\usepackage{graphicx}
\usepackage{amsmath,amssymb}
\usepackage{amsthm}
\usepackage{enumerate}
\usepackage{dsfont}
\usepackage{subfigure}
\usepackage[dvips]{color}
\usepackage{wasysym}
\usepackage{bbold}
\newcommand\bea{\begin{eqnarray}}
\newcommand\eea{\end{eqnarray}}
\newcommand\beq{\begin{equation}}
\newcommand\eeq{\end{equation}}
\newcommand\scalemath[2]{\scalebox{#1}{\mbox{\ensuremath{\displaystyle #2}}}}

\newcommand{\noi}{\noindent}
\newcommand{\bib}{\bibitem}

\def\nn{\nonumber}
\def\f{\frac}
\def\al{\alpha}

\def\de{\delta}

\def\ga{\gamma}
\def\Ga{\Gamma}
\def\si{\sigma}
\def\Do{\partial}
\def\De{\Delta}
\def\dg{\dagger}

\def\ua{\uparrow}
\def\da{\downarrow}

\def\be{\beta}

\def\th{\theta}
\def\one{\mathds{1}}

\def\inf{\infty}
\def\sq{\sqrt}
\def\ot{\otimes}

\begin{document}

\title{\bf{Transport across a junction of topological insulators and
a superconductor}}

\author{Abhiram Soori$^1$, Oindrila Deb$^1$, K. Sengupta$^2$
and Diptiman Sen$^1$}
\affiliation{$^1$ Centre for High Energy Physics, Indian Institute of Science,
Bangalore 560 012, India \\
$^2$ Theoretical Physics Department, Indian Association for the Cultivation
of Science, Jadavpur, Kolkata 700 032, India}

\begin{abstract}
We study transport across a line junction lying between two
orthogonal topological insulator surfaces and a superconductor which
can have either $s$-wave (spin-singlet) or $p$-wave (spin-triplet)
pairing symmetry. We present a formalism for studying the effect of
a general time-reversal invariant barrier at the junction and show
that such a barrier can be completely described by three arbitrary
parameters. We compute the charge and the spin conductance across
such a junction and study their behaviors as a function of the bias
voltage applied across the junction
and the three parameters used to characterize the barrier.
We find that the presence of topological insulators and a
superconductor leads to both Dirac and Schr\"odinger-like features
in charge and spin conductances. We discuss the effect of bound
states on the superconducting side of the barrier on the
conductance; in particular, we show that for triplet $p$-wave
superconductors such a junction may be used to determine the spin
state of its Cooper pairs. Our study reveals that there is a non-zero
spin conductance for some particular spin states of the triplet Cooper 
pairs; this is an effect of the topological insulators
which break the spin rotation symmetry. Finally, we find an unusual 
satellite peak (in addition to the usual zero bias peak) in the spin 
conductance for $p$-wave symmetry of the superconductor order parameter.
\end{abstract}

\maketitle

\section{Introduction}~\label{sec-intro}
Recent theoretical~\cite{zhang1,kane1,kane2,teo,qi} and
experimental~\cite{hasan1,exp2,exp1,hasan2} works have led to the
discovery of a new class of materials called topological
insulators~(TI). In these materials 
the surface states have a gapless spectrum governed by a massless Dirac
equation~\cite{ti_rev}; these states contribute to charge transport at low
temperatures. Such materials could be either in two or three spatial
dimensions. A two-dimensional TI will host a one-dimensional gapless
edge states while three-dimensional~(3D) TI will host gapless states
protected by time-reversal symmetry on its two-dimensional surface.
The 3D~TIs can be classified as strong or weak depending on whether
the number of Dirac cones is odd or even, and this number is
determined by a topological invariant. The odd number of Dirac cones
on the surface of a strong TI is protected against time-reversal
invariant (for instance, non-magnetic) perturbations for topological
reasons \cite{kane2,zhang1,hasan1}. For materials such as $\rm Bi_2
Te_3$ and ${\rm Bi_2 Se_3}$, surfaces have been found which have a
single Dirac cone near the $\Gamma$ point of the 2D surface
Brillouin zone~\cite{hasan1,exp2,hasan2}. Many interesting features
of the surface Dirac electrons have been
studied~\cite{kane4,fu,been1,tanaka1,mondal1,hasan2,
burkov,garate,zyuzin,das1}. Some of these studies involve interfaces
created in a TI using proximate magnetic or superconducting
materials or gate
voltages~\cite{kane4,been1,tanaka1,burkov,garate,das1,rein}.
Junctions of different surfaces of a TI (in some cases separated by
a geometrical step or a magnetic domain
wall)~\cite{taka,sen12,wickles,biswas,alos,sitte,zhang3,apalkov} or
of surfaces of a TI with normal metals or magnetic
materials~\cite{modak} have also been studied. However, junctions of
multiple TI surfaces with a superconductor have not been studied in
detail so far; this is what we aim to do in this paper.

The problem of electron transmission across a junction of a normal
metal~(NM) and a superconductor (SC) has been extensively studied
for many years~\cite{btk,kasta,hayat,tanaka95,kseng01,kseng02,kwon}.
It is well-known that the sub-gap transport in such junctions, for
small barrier strengths, is governed by Andreev reflection while for
large barrier strengths they reflect the sub-gap quasiparticle
density of states (DOS)\cite{btk,tanaka1}. Since the presence of
localized quasiparticle edge states below the gap depends on the order
parameter symmetry, such transport measurements, in the strong
barrier limit, provide us with a tool for determining the pairing
symmetry of the superconductor. In contrast, in the weak barrier
limit, it was demonstrated that the conductance of a NM-SC junction
can be more than the NM-NM conductance owing to Andreev reflection
\cite{kasta}. More recently, hybrid
superconductor-semiconductor~(${\rm Bi_2Te_3}$) devices has been
successfully fabricated and conductance measurements in such systems
has been carried out~\cite{hayat}. We note that these systems are
somewhat akin to the junctions that we study in this work. However,
the junction of a TI and a SC is expected to be more complex than
its NM-SC counterpart. The reason for this complexity arises out of
the fact that two-dimensional~(2D) surface states of a TI display
spin-momentum locking; hence scattering at a junction that changes
the electron momentum couples different components of the electron
spin. In contrast, the particle and hole are coupled to each other
on the SC side. Hence a treatment of the transport between a TI and
a SC necessarily requires us to use a four-component spinor
formalism describing both spin and particle-hole degrees of freedom
\cite{kseng01,kseng02,kwon}.

\begin{figure}[htb]
\epsfig{figure=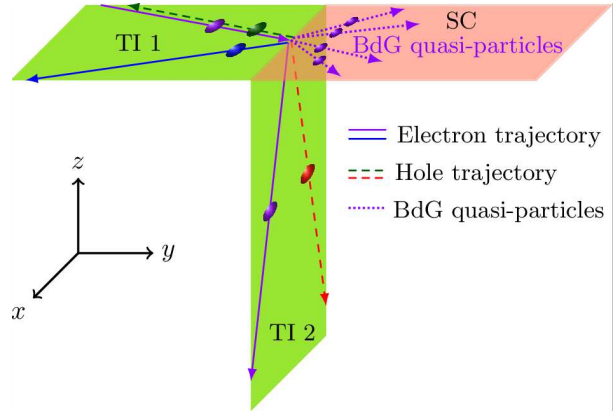,width=8.0cm}
\caption{Schematic diagram of the proposed system with a line junction.
Topological insulator surfaces 1 (lying in the $x-y$
plane with $y<0$) and 2 (lying in the $x-z$ plane with $z<0$) and a
superconductor (lying in the $x-y$ plane with $y>0$) meet at a junction given
by the line $y=z=0$. We indicate several processes which can be triggered by
an electron incident from the TI-1 side: normal or Andreev reflection back
to TI-1, normal of Andreev transmission to the TI-2 sides, and four possible
BdG quasiparticle transmissions on the SC side (which eventually decay to
zero and get converted to Cooper pairs).} \label{fig_schem} \end{figure}

In this work, we will analyze the transport properties for the
system shown in Fig.\ \ref{fig_schem}. The system consists of two
orthogonal TI surfaces, called TI-1 and TI-2. There is a
two-dimensional SC surface which can be formed experimentally by
depositing a 2D superconducting film on the surface of a 3D
insulator. In what follows, we shall consider both singlet $s$-wave
and triplet $p$-wave pairing symmetries for the superconducting
film. The two TIs and the SC are separated by a line junction. We
will consider a Dirac electron incident on the line junction from
TI-1 with arbitrary angle of incidence and study its reflection
(normal and Andreev) back to TI-1 and its transmission (normal and
Andreev) into the TI-2 and the superconducting film. The main
results that we obtain from such an analysis are the following.
First, we develop appropriate boundary conditions for studying
transmission across such a junction involving Dirac electrons in the
TI and Schr\"odinger electrons in the superconductor. The general
time-reversal invariant boundary condition is found to involve three
parameters which can be interpreted as the strengths of three
barriers close to the junction on the TI-1, TI-2 and SC sides.
Second, using these boundary conditions, we compute the charge and
spin conductances of such a junction as functions of the barrier
strengths and the bias voltage and thus compare and contrast the
properties of sub-gap transport in these junctions with their
conventional counterparts. In particular, we find that for
$\chi_1 = \chi_2 = 0$, these junctions never reach the maximum charge
conductance value $2G_0$ (where $G_0$ is an unit of conductance defined in
Eq.~\eqref{g123}) found in conventional NM-SC junctions.
Third, we find that, in contrast to all conventional NM-SC junctions, TI-SC
junctions can be used to distinguish between the different spin
states of the Cooper pair of a triplet $p$-wave superconductor. We show
that such a property stems from the spin-momentum locking of the
electrons on the TI surfaces. Fourth, we find that for a $p$-wave
SC, the differential conductance has a zero-bias peak
similar to the conventional NM-SC junctions; however for finite
biases, the conductance both oscillates and decreases as the barrier
strength increases which is to be contrasted with the monotonically
decreasing nature of sub-gap conductance in conventional NM-SC
junctions. Thus these junctions display both Dirac-like and
Schr\"odinger-like characters. Finally, we find that there is a 
non-zero spin conductance for one particular component of the spin
and two possible spin states of the triplet Cooper pairs. Further, 
we find that the differential spin conductance displays an unusual 
satellite peak away from zero bias and study the behavior of this peak 
with chemical potentials of the TI and SC surfaces and the barrier
strengths of the junction.

The detailed plan of this paper is as follows. In Sec.~\ref{sec-ham}
we write down the Hamiltonian for electrons on the topological
insulator surfaces and the superconductors for both $s-$ and
$p-$wave pairings and discuss their basic properties which will be
useful in subsequent analysis. This is followed by
Sec.~\ref{sec-curr} where we chart out the boundary conditions
appropriate for transport through the junction. We utilize current
conservation to discuss how the current gets converted from single
quasiparticles (electrons or holes) near the junction to Cooper
pairs deep inside the SC. Next, in Sec.~\ref{sec-cond}, we use the
boundary conditions to compute the transmission amplitudes of
electrons from the TI-1 into the TI-2 and the SC and hence the
charge conductance of the junction for arbitrary bias voltage,
chemical potential difference, and barrier potential parameters. We
discuss the obtained results in details in Sec.\ \ref{sec-res}. This
is followed by Sec.~\ref{sec-spcurr} where we study the spin
transport through a $p$-wave SC with different pairing symmetries.
Finally, in Sec.\ \ref{sec-disc}, we present a discussion of our
results and some possible experiments to test our theory.

\section{Hamiltonians}~\label{sec-ham}
The Hamiltonian for the electrons on the surface of a TI is given by
\cite{mondal1} \beq H_{\hat{n}} = \int \f{dk_i dk_j}{(2\pi)^2}
\psi^{\dg}_{\vec k} (\hbar v_F \hat{n} \cdot \vec \si \times \vec
k-\mu_{\hat n} \one) \psi_{\vec k}, \eeq where $v_F$ is the Fermi
velocity, the unit vector $\hat n$ points normal to the 2D surface,
$k_i,~k_j$ are momenta in the 2D plane, and $\psi_{\vec k}
=(\psi_{\vec k,\ua}, ~\psi_{\vec k,\da})^T$ is a two-component
spinor. For TI-1, $\hat n= \hat z$ and $(i,j)=(x,y)$, while for
TI-2, $\hat n=\hat y$ and $(i,j)=(x,z)$. The energy-momentum
dispersion and the eigenstates on the TI-1 side are given by
\bea E &=& -\mu_1\pm \hbar v_1\sq{k_x^2+k_y^2}, \nn \\
\psi_1^{\pm} &=& (1/\sq{2}) ~[1, ~\mp ie^{i\th_1}]^T , \nn \\
e^{i\th_1} &=& (k_x+ik_y)/\sq{k_x^2+k_y^2}, \label{disp1}\eea
where $v_1$ is the Fermi velocity, and $\th_1$ lies in the range $[0, \pi]$.
In our calculations, we will consider the band which has the $+$ sign in the
expression for the energy by choosing $\mu_1>0$ and working at energies
$|E|<\mu_1$. On the TI-2 side, the dispersion and the eigenstates are given by
\bea E &=& -\mu_2 + \hbar v_2\sq{k_x^2+k_z^2}, \nn \\
\psi_2^+ &=& (1/\sq{2}) ~[\sq{1+\cos{\th_2}}, ~\sq{1- \cos{\th_2}}]^T , \nn \\
e^{i\th_2} &=& (k_x-ik_z)/\sq{k_x^2+k_z^2}~, \label{disp2}\eea
where $v_2$ is the Fermi velocity. To keep the discussion simple we choose
$\mu_1=\mu_2=\mu_{TI}$. Also, our numerical results will be obtained for the
case $v_1=v_2$ although we have retained the general form ($v_2\le v_1$) in
the important analytical expressions.

The wave functions on the SC side are described by a four-component
Bogoliubov-de Gennes (BdG) spinor $\Psi$ whose upper two components
correspond to particles and lower two components correspond to
holes. Namely, $\Psi =
[\psi_{\ua},~\psi_{\da},~\psi^*_{\ua},~\psi^*_{\da}~]^T$. (Note that
many papers use the Nambu convention, with $\Psi =
[\psi_{\ua},~\psi_{\da},~-\psi^*_{\da},~\psi^*_{\ua}~]^T$. In that
convention, for example, factors of $i\si^y$ will not appear in
Eq.~\eqref{pair-s}).
In terms of the Pauli matrices $\si^{x,y,z}$ and $\tau^{x,y,z}$ which act on 
the spin and particle-hole components respectively, the Hamiltonian on the SC
side can be written as 
\bea H_3 ~=~ \int_{-\inf}^{\inf} dx\int_{0}^{\inf}dy~ \Psi^{\dg} (x,y)
\Big[ && \Big(-\f{\hbar^2 \vec\nabla^2}{2m} -\mu_{SC} \Big)\tau^z \nn \\
&&+ ~\De(x,y)\Big]\Psi (x,y), \nn \\
&& \label{HSC} \eea
where $\De(x,y)$ is called the pair potential. For our analysis of an
$s$-wave SC (in which Cooper pairs form a spin singlet), we will take
the pair potential to be of the form
\bea \De(x,y) &=& \left( \begin{array}{cc}
0 & \De_0 ~i\si^y\\
-\De_0 ~i\si^y & 0
\end{array} \right). \label{pair-s} \eea {\bea \De(x,y) &=&
\left(
\begin{array}{cc}
0 & \De_0 f(\vec k) (\vec d \cdot \vec \si) i\si^y\\
-\De_0 f^*(\vec k) (\vec d \cdot\vec \si^*) i\si^y & 0
\end{array} \right) \nn \\
~~&&~~\label{pair-p} \eea
where $\vec d$ is a unit vector with real components, and $f (\vec k)$
is defined below.
Physically, $\vec d$ governs the spin pairing of the Cooper pairs
in the $p$-wave SC. For $\vec d = \hat x$, $\hat y$ and $\hat z$, a
consideration of the matrix $(\vec d \cdot \vec \si) i\si^y$ shows that the
Cooper pairs are in the spin states $\ua \ua - \da \da$, $\ua \ua + \da \da$,
and $\ua \da + \da \ua$, respectively. (In principle, $\vec d$ could depend
on the momentum $\vec k$. However, in this work we are going to study
systems with constant $\vec d$). In Eq.~\eqref{pair-p}, $f(\vec k)$ is
a dimensionless linear function of $\vec k$ defined as
\bea f(\vec k) &\equiv& -i (a_x \Do_x + a_y \Do_y), \nn \\
f^*(\vec k) &\equiv& i (a^*_x \Do_x + a^*_y \Do_y), \eea
where $a_x, a_y$ are constants which may be complex. [In this paper we will
consider wave functions in the SC which are of the form $e^{i(k_x x + k_y y)}$
where $k_x$ is real but $k_y$ may be complex. Hence there will not be any
simple relation between $f(\vec k) = \vec a \cdot \vec k$ and $f^*(\vec k)
= - \vec a^* \cdot \vec k$ in general]. In both Eqs.~(\ref{pair-s}) and
(\ref{pair-p}), $\Delta_0$ will be assumed to be a real parameter with
dimensions of energy.

In a SC, the nature of the energy-momentum dispersion depends crucially on
the symmetry of the superconducting pair potential. The dispersion is given by
\beq E(\vec k) = \pm \sq{\Big[\f{\hbar^2(k_x^2+k_y^2)}{2m}- \mu_{SC}\Big]^2
+ \De_0^2} \label{disp3s} \eeq
for an $s$-wave SC. For a $p$-wave SC, we can use the fact that $(\vec d \cdot
\vec \si) i\si^y (\vec d \cdot \vec \si^*) i\si^y = \one$ to show that
\beq E = \pm \sq{\Big[\f{\hbar^2(k_x^2+k_y^2)}{2m}- \mu_{SC}\Big]^2
+ \De_0^2 ~(\vec a \cdot \vec k)(\vec a^* \cdot \vec k)}.
\label{disp3p} \eeq The important difference to note between the
dispersions for these two kinds of pair potentials is that the
superconducting gap is always isotropic for $s$-wave symmetry while
in the case of $p$-wave, it is isotropic only when $(\vec a \cdot
\vec k)(\vec a^* \cdot \vec k)$ is invariant under rotation in $x-y$ plane.

\section{Probability and Charge Currents and Boundary
Conditions}~\label{sec-curr}
In this section, we will first discuss
the equations of motion. Then we will define a probability density
$\rho_p$ and current ${\vec J}_p$, and a charge density $\rho_c$ and
current ${\vec J}_c$. It is well known that the corresponding
continuity condition for superconductors necessitates conversion
between particle and condensate currents as discussed for $s$-wave in
Ref.\ \onlinecite{btk}. Here we shall discuss the continuity
condition for both $s$- and $p$-wave superconductors using the four
component formulation.

Given the two-component spinor $\psi$, let us define $\phi = \psi^*$.
For a TI with a unit normal $\vec n$, the equations of motion for $\psi$ and
$\phi$ are given by
\bea i \hbar \Do_t \psi &=& [\hbar v_F {\hat n} \cdot {\vec \si} \times
(-i {\vec \nabla}) ~-~ \mu_{TI}] ~\psi, \nn \\
i \hbar \Do_t \phi &=& [\hbar v_F {\hat n} \cdot {\vec \si}^* \times
(-i {\vec \nabla}) ~+~ \mu_{TI}] ~\phi. \label{eomti} \eea
For an $s$-wave SC, the equations of motion are
\bea i \hbar \Do_t \psi &=& \left(- \f{\hbar^2 {\vec \nabla}^2}{2m} ~-~
\mu_{SC} \right) \psi ~+~ \Delta_0 i \si^y \phi, \nn \\
i \hbar \Do_t \phi &=& -~ \left(- \f{\hbar^2 {\vec \nabla}^2}{2m} ~-~
\mu_{SC} \right) \phi ~-~ \Delta_0 i \si^y \psi. \label{eoms} \eea
For a $p$-wave SC, the equations of motion are
\bea i \hbar \Do_t \psi &=& \left(- \f{\hbar^2 {\vec \nabla}^2}{2m} ~-~
\mu_{SC} \right) \psi \nn \\
& & +~ \Delta_0 f(\vec k) ({\vec d} \cdot {\vec \si}) i \si^y \phi,
\nn \\
i \hbar \Do_t \phi &=& -~ \left(- \f{\hbar^2 {\vec \nabla}^2}{2m} ~-~
\mu_{SC} \right) \phi \nn \\
& & -~ \Delta_0 f^*(\vec k) ({\vec d} \cdot {\vec \si}^*) i \si^y
\psi. \label{eomp} \eea

In deriving the equations of motion for $\phi$ in
Eqs.~(\ref{eomti}-\ref{eomp}),
we have used the fact that $\phi$ is the complex conjugate of $\psi$. We will
now begin to treat $\psi$ and $\phi$ as independent variables which are {\it
not} related by complex conjugation; for example, when solving the
equations of motion, we will assume that $\psi$ and $\phi$ depend on the 
time $t$ in exactly the same way, i.e., as $e^{-iEt/\hbar}$.
The upper two and lower two components of the BdG
spinor $\Psi$ will be given by $\psi$ and $\phi$ respectively.

The probability and charge densities are given by
\bea \rho_p &=& \psi^\dg \psi ~+~ \phi^\dg \phi ~=~ \Psi^\dg \Psi, \nn \\
\rho_c &=& e(\psi^\dg \psi ~-~ \phi^\dg \phi) ~=~ e\Psi^\dg \tau^z \Psi. \eea
We will now look for currents ${\vec J}_p$ and ${\vec J}_c$ which can satisfy
the continuity equations of continuity, namely, $\Do_t \rho_p + {\vec \nabla}
\cdot {\vec J}_p = 0$ and $\Do_t \rho_c + {\vec \nabla} \cdot {\vec J}_c = 0$.
In the TIs, we find that the currents
\bea {\vec J}_p &=& v_F [\psi^\dg {\hat n} \times {\vec \si} \psi ~+~ \phi^\dg
{\hat n} \times {\vec \si}^* \phi], \label{jpti} \\
{\vec J}_c &=& e v_F [\psi^\dg {\hat n} \times {\vec \si} \psi ~-~ \phi^\dg
{\hat n} \times {\vec \si}^* \phi], \label{jcti} \eea
satisfy the respective continuity equations.

\noi {\bf $s$-wave SC:}
In an $s$-wave SC governed by Eq.~(\ref{pair-s}), we find that
\bea {\vec J}_p &=& \f{\hbar}{m} Im (\psi^\dg {\vec \nabla} \psi) ~-~
\f{\hbar}{m} Im (\phi^\dg {\vec \nabla} \phi) \nn \\
&=& \f{\hbar}{m} Im (\Psi^\dg \tau^z {\vec \nabla} \Psi) \label{jps} \eea
(where $Im$ denotes the imaginary part) satisfies the continuity equation
with $\rho_p$. However, it is well known that
\bea {\vec J}_c &=& \f{e\hbar}{m} Im (\psi^\dg {\vec \nabla} \psi) ~+~
\f{e\hbar}{m} Im (\phi^\dg {\vec \nabla} \phi) \nn \\
&=& \f{e\hbar}{m} Im (\Psi^\dg {\vec \nabla} \Psi) \label{jcs} \eea
does not satisfy the continuity equation with $\rho_c$. Instead,
if we define the Cooper pair current~\cite{btk}
\bea {\vec J}_{pair} &=& - ~{\hat y} ~\f{4 e\Delta_0}{\hbar} ~\int_0^y dy' ~
Re [\psi^\dg (x,y') \si^y \phi (x,y')], \nn \\
&& \label{jpair}\eea
then the total charge current in the superconductor, defined as
${\vec J}_{tot} = {\vec J}_c + {\vec J}_{pair}$, satisfies $\Do_t \rho_c +
{\vec \nabla} \cdot {\vec J}_{tot} = 0$. For a state which has definite values
of the energy $E$ and the momentum $k_x$, we have $\Do_t \rho_c =0$ and
$\Do_x J_{x,tot} = 0$; hence $J_{y,tot} = J_{y,c} + J_{y,pair}$ will also be
independent of the $y$ coordinate. However $J_{y,c}$ and $J_{y,pair}$ will
separately vary with $y$; as $y$ goes from 0 to $\infty$, $J_{y,c}$ will go
from a finite value to 0 while $J_{y,pair}$ will go from 0 to a finite value.
In other words, the single electron or hole current $J_{y,c}$ will gradually
get converted to the Cooper pair current $J_{y,pair}$ as $y$
increases~\cite{btk}.

\noi {\bf $p$-wave SC:}
In a $p$-wave SC governed by Eq.~(\ref{pair-p}), we find that if $f(\vec k) =
-i (a_x \Do_x + a_y \Do_y)$, then
\bea {\vec J}_p &=& \f{\hbar}{m} Im (\Psi^\dg \tau^z {\vec \nabla} \Psi) \nn \\
&& + ~{\hat x} ~\f{2 \Delta_0}{\hbar} ~Re [ a_x \psi^\dg (\vec d \cdot
\vec \si) i \si^y \phi] \nn \\
&& + ~{\hat y} ~\f{2 \Delta_0}{\hbar} ~Re [ a_y \psi^\dg (\vec d \cdot
\vec \si) i \si^y \phi] \label{jpp} \eea
satisfies the continuity equation with $\rho_p$. (Note that the expression for
${\vec J}_p$ in the $p$-wave case contains a term proportional to $\Delta_0$,
unlike the expression in the $s$-wave case which does not have such a term).
The charge current ${\vec
J}_c$ defined in Eq.~(\ref{jcs}) again fails to satisfy the continuity
equation with $\rho_c$. But if we define the Cooper pair current
\bea {\vec J}_{pair} &=& -~ {\hat y} ~\f{2 e \Delta_0}{\hbar} \nn \\
&\times& \int_0^y dy' ~Re [ ~\psi^\dg (x,y') (\vec d \cdot \vec \si) \si^y
f (\vec k) \phi (x,y') \nn \\
&& ~~~~~~~~~~~~- (f (\vec k) \psi^\dg (x,y')) (\vec d \cdot
\vec \si) \si^y \phi (x,y')]. \nn \\
&& \eea
the total current
${\vec J}_{tot} = {\vec J}_c + {\vec J}_{pair}$ satisfies $\Do_t \rho_c +
{\vec \nabla} \cdot {\vec J}_{tot} = 0$.

\section*{Boundary Conditions}
We now turn to the boundary conditions which need to be imposed at
the junction to ensure that the component of the probability current
normal to the junction is conserved. We will denote the currents on
the three sides of the junction by ${\vec J}_{p1}$, ${\vec J}_{p2}$,
and ${\vec J}_{p3}$. On the TI-1 and TI-2 sides, the incoming
currents are given by $({\hat y} \cdot {\vec J}_{p1})_{y \to 0-}$
and $({\hat z} \cdot {\vec J}_{p2})_{z \to 0-}$ respectively, while
on the SC side, the outgoing current is $({\hat y} \cdot {\vec
J}_{p3})_{y \to 0+}$. The current conservation condition is
therefore \beq ({\hat y} \cdot {\vec J}_{p1})_{y \to 0-} ~-~ ({\hat
z} \cdot {\vec J}_{p2})_{z \to 0-} ~=~ ({\hat y} \cdot {\vec
J}_{p3})_{y \to 0+}. \label{curcon} \eeq Using the expressions for
${\vec J}_p$ given in Eqs.~(\ref{jpti}), (\ref{jps}) and
(\ref{jpp}), we find that Eq.~(\ref{curcon}) implies \beq v_1
\Psi_1^\dg \si^x \Psi_1 ~-~ v_2 \Psi_2^\dg \si^x \Psi_2 ~=~
\f{\hbar}{m} ~Im (\Psi_3^\dg \tau^z \Do_y \Psi_3) \label{curcons}
\eeq if the SC is $s$-wave, and
\bea && v_1 \Psi_1^\dg \si^x \Psi_1 ~-~ v_2 \Psi_2^\dg \si^x \Psi_2 \nn \\
&=& \f{\hbar}{m} ~Im (\Psi_3^\dg \tau^z \Do_y \Psi_3) ~+~\f{2 \Delta_0}{\hbar}~
Re [ a_y \psi_3^\dg (\vec d \cdot \vec \si) i \si^y \phi_3]. \nn \\
&& \label{curconp} \eea
if the SC is $p$-wave. Now onwards, we shall use the four-component spinor
$\Psi$ instead of the two-component spinors $\psi$ and $\phi$.

To find the general boundary condition at the junction which satisfies
Eqs.~(\ref{curcons}) and (\ref{curconp}), let us assume that there are
three barriers which lie on the TI-1, TI-2 and SC sides of the junction
and are located very close to the junction; we will model them as
$\de$-function barriers with dimensionless strengths $\chi_1$, $\chi_2$
and $\chi_3$ respectively. We will assume that these barriers are
invariant under time reversal, for instance, that they do not involve
any magnetic fields. Then an analysis similar to that in
Ref.~\onlinecite{modak} will give the following boundary conditions at
the junction for the case of an $s$-wave SC,
\bea \Psi_3 &=& c ~\big[M(\chi_1) \Psi_1 +
\be M^{\dg}(\chi_2) \Psi_2 \big], \nn \\
\f{\hbar}{m v_1} \Do_y \Psi_3 &-& 2\chi_3\Psi_3 \nn \\
&=&\f{i}{c}\si^x\ot\tau^z ~\big[ M(\chi_1) \Psi_1 -
\be M^{\dg}(\chi_2) \Psi_2 \big], \nn \\
{\rm where~}& M(\chi)&=\cos{\chi} - i \sin{\chi} \si^x \ot \tau^z
\label{bc-s} \eea
where $\be$ is related to the ratio of the Fermi velocities in TI-1 and
TI-2 as $\be= \sqrt{v_2/v_1}$.

For the case of a $p$-wave SC, a simple generalization of the boundary
condition in Eqs.~(\ref{bc-s}) which conserves the probability current at
the junction is given by
\bea \Psi_3 &=& c \big[ M(\chi_1) \Psi_1 +
\be M^{\dg}(\chi_2) \Psi_2 \big], \nn \\
\f{\hbar}{m v_1} \Do_y \Psi_3 &-& 2\chi_3 \Psi_3 \nn \\
&+&\f{\Delta_0}{\hbar v_1} \left( \begin{array}{cc}
0 & -a_y ({\vec d} \cdot {\vec \si}) \si^y \\
a^*_y ({\vec d} \cdot {\vec \si}^*) \si^y & 0 \end{array} \right) \Psi_3
\nn \\
&=&\f{i}{c}\si^x\ot\tau^z ~\big[ M(\chi_1) \Psi_1 -
\be M^{\dg}(\chi_2) \Psi_2 \big]. ~~~~
\label{bc-p} \eea
It can be shown by a simple calculation that the boundary conditions in
Eqs.~(\ref{bc-s}-\ref{bc-p}) also satisfy charge current conservation
at the junction.

Note that Eqs.~(\ref{bc-s}-\ref{bc-p}) contain a real dimensionless parameter
$c$. A precise determination of the value of $c$ requires a microscopic
knowledge of the junction and is beyond the scope of the present work.
A similar parameter appears in the study of junctions in other
systems~\cite{raoux,carreau}. In the limits $c \to 0$ or $\infty$, the SC gets
decoupled from the TIs and the system reduces to one which only involves two
TI surfaces. (The problem of two TIs with a junction has been studied in
Ref.~\onlinecite{sen12}). In this paper we will set $c=1$ in all our
numerical calculations.

\section{Conductance Calculations}~\label{sec-cond}
We are interested in the charge transport at a sub-gap applied
voltage between the TIs and the SC. Our calculation will proceed as
follows. Given an electron incident on the junction from the TI-1
side with unit amplitude at an angle of incidence $\th_1$ and energy
$E$ and using the boundary conditions (Eqs.\ \ref{bc-s} and
\ref{bc-p}), we shall compute the amplitudes of eight other wave
functions, namely, normally reflected electrons and Andreev
reflected holes on the TI-1 side (with amplitudes $r_N$ and $r_A$
respectively), normally transmitted electrons and Andreev
transmitted holes on the TI-2 side (with amplitudes $t_N$ and
$t_A$), and four electron and hole-like BdG quasiparticle wave
functions on the SC side with amplitudes $t_1$, $t_2$, $t_3$ and
$t_4$. Note that in what follows we shall set the zero energy at the
middle of the superconducting energy gap.

\subsection{Wave Functions in the TIs}
To account for a possible Andreev reflection and Andreev transmission in TI-1
and TI-2, we have to write down the wave functions in the TIs as
four-component BdG spinors. Due to translational invariance along $\hat x$,
all the wave functions will have a factor of $e^{ik_xx}$. Also, all the
excitations will be taken to have an energy
$E$ which means that there will be a common factor of $e^{-iEt/\hbar}$.
With $r_N$ and $r_A$ being the amplitudes for normal and Andreev reflections,
we can write the wave function in TI-1 as
\bea \Psi_1 &=& \f{1}{\sq{2}} \left( \begin{array}{c}
1 \\ -ie^{i\th_1} \\ 0 \\ 0
\end{array} \right)e^{ik_yy}
+ \f{r_N}{\sq{2}} \left( \begin{array}{c}
1 \\ -ie^{-i\th_1} \\ 0 \\ 0
\end{array} \right)e^{-ik_yy} \nn \\
&& + \f{r_A}{\sq{2}} \left( \begin{array}{c}
0 \\ 0 \\ 1 \\ -ie^{-i\th_{1h}}
\end{array} \right)e^{ik_{yh}y}. \label{psi-1} \eea
Here, $(k_x,k_y)$ is the momentum of the incident electron on TI-1 which is
related to $(E,\th_1)$ as in Eq.~(\ref{disp1}). The normally
reflected electron has a momentum $(k_x,-k_y)$ and the Andreev reflected hole
has momentum $(k_x,k_{yh})$ where $k_{yh}=\sq{(\mu_{TI}-E)^2/(\hbar v_1)^2
-k_x^2}$ and $e^{i\th_{1h}}=(k_x+ik_{yh})/\sq{k_x^2+k_{yh}^2}$. For the case
$E=0$, the hole will have a momentum $(k_x,k_y)$ just like the incident
electron, but its group velocity ${\vec v}_g = {\vec \nabla}_{\vec k} E$
will be opposite to the incident electron's group velocity. This is called
retroreflection. At a given energy $E>0$, since $\nu_E \equiv (\mu_{TI}+E)/
(\mu_{TI} -E)>1$,
\beq k_{yh} ~=~ \frac{\mu_{TI}-E}{\hbar v_1} ~\sqrt{1 ~-~ \nu_E^2 \cos^2
\th_1} \eeq
becomes purely imaginary with $Im(k_{yh})<0$ for
a certain range of $\th_1$. This corresponds to an evanescent
Andreev mode. Such modes exist for the ranges $0\le\th_1<\th_{1E}$ and
$\pi - \th_{1E} < \th_1 \le \pi$, where $\th_{1E}=\cos^{-1}({1/\nu_E})$.

On the TI-2 side we may have either a normally transmitted electron with
momentum $(k_x,-k_z)$ or an Andreev transmitted hole with momentum
$(k_x,k_{zh})$. The longitudinal momenta for the transmitted
electron and hole are $k_z=\sq{(\mu_{TI}+E)^2/(\hbar v_2)^2-k_x^2}$ and
$k_{zh}=\sq{(\mu_{TI}-E)^2/(\hbar v_2)^2-k_x^2}$ respectively. [For $E=0$,
the momentum of the hole will be $(k_x,k_z)$].
At a given energy $E$, when $\nu_E>v_1/v_2$, there exist evanescent
Andreev modes~($Im(k_{zh})<0$) for the following ranges of the incident
angle $\th_1$: $0\le\th_1<\th_{2E}$ and $\pi - \th_{2E} < \th_1 \le \pi$,
where $\th_{2E}=\cos^{-1}[{v_1/(v_2\nu_E)}]$. With $t_N$ and $t_A$
being the amplitudes for normal and Andreev transmissions, we can
write the wave function in TI-2 as
\bea \Psi_2 &=& \f{t_N}{\sq{2}} \left( \begin{array}{c}
\sq{1+\cos{\th_2}} \\ \sq{1-\cos{\th_2}} \\ 0 \\ 0
\end{array} \right)e^{-ik_z z} \nn \\
&& +~ \f{t_A}{\sq{2}} \left( \begin{array}{c}
0 \\ 0 \\ \sq{1-\cos{\th_{2h}}} \\ \sq{1+\cos{\th_{2h}}}
\end{array} \right)e^{ik_{zh} z}, \nn \\
{\rm and} && e^{i\th_2}=(k_x+ik_z)/\sq{k_x^2+k_z^2},\nn \\
&&~~e^{i\th_{2h}}=(k_x+ik_{zh})/\sq{k_x^2+k_{zh}^2}.
\label{psi-2} \eea

\subsection{Wave Function for $s$-wave SC}~\label{sec-swave}
For the case of an $s$-wave SC, as mentioned earlier the pair potential is
isotropic and $\vec k$ independent as shown in Eq.~(\ref{pair-s}).
The momentum along $\hat x$ will be equal to $k_x$ (the same
as in the TIs) due to the translational invariance of the system in the
$\hat x$ direction, while the longitudinal momentum $k_{ySC}$ will be given by
\beq k_{ySC} = \pm k_F\sq{1 -\f{k_x^2}{k_F^2}
\pm i \sq{\f{\De_0^2-E^2}{\mu_{SC}^2}}}~, \label{kySC-s}\eeq
where we choose the $\pm$ signs in such a way that $Im (k_{ySC}) > 0$.
Of the four possible solutions for $k_{ySC}$ in the above equation, we must
choose the two for which the wave functions decay as $y\to\inf$; they
can be written in the form $k_{ySC}=\pm k_R+ik_I$ where $k_R$ and $k_I$
are positive. At energies in the gap~($|E|<\De_0$), the wave function on
SC looks like
\bea \Psi_3(y) &=& \Big[ \f{t_1}{\sq{2}} \left( \begin{array}{c}
1 \\ 0 \\ 0 \\ e^{i\eta}
\end{array} \right) + \f{t_2}{\sq{2}} \left( \begin{array}{c}
0 \\ 1 \\ -e^{i\eta} \\ 0
\end{array} \right) \Big] e^{(ik_R-k_I) y} \nn \\
&+& \Big[ \f{t_3}{\sq{2}} \left( \begin{array}{c}
1 \\ 0 \\ 0 \\ e^{-i\eta}
\end{array} \right) + \f{t_4}{\sq{2}} \left( \begin{array}{c}
0 \\ 1 \\ -e^{-i\eta} \\ 0
\end{array} \right) \Big] e^{(-ik_R-k_I) y} \nn \\
&& {\rm where} ~~~e^{i\eta} ~=~ \f{1}{\Delta_0} ~(E-i\sq{\De_0^2-E^2}).
\label{psi-3s} \eea

\subsection{Wave Function for $p_y$-wave SC and
$\vec d=\hat z$}~\label{sec-pwave}
For a $p$-wave SC with $f(\vec k)=k_y/k_F$ (where $\hbar k_F =
\sqrt{2m \mu_{SC}}$) and $\vec d=\hat z$, the dispersion is anisotropic.
For a given $E$ and $\th_1$ on TI-1, the longitudinal momentum on the SC
side $(\pm k_R + ik_I)$ is given by
\bea k_{ySC} &=& \pm k_F\sq{1-\Ga_1 \pm i \sq{\f{\De_0^2-E^2}{\mu_{SC}^2} -
\Ga_2}}, \nn \\
{\rm~where~} \Ga_1 &=& \f{k_x^2}{k_F^2} + \f{\De_0^2}{2\mu_{SC}^2}, \nn \\
\Ga_2 &=& \f{\De_0^2}{\mu_{SC}^2}\f{k_x^2}{k_F^2} +
\f{\De_0^4}{4\mu_{SC}^4},
\label{kySC-py-dz} \eea
and we choose the $\pm$ signs so that $Im (k_{ySC}) > 0$. The wave function
is given by
\bea \Psi_3(y) &=& \Big[ \f{t_1}{\sq{2}} \left( \begin{array}{c}
1 \\ 0 \\ 0 \\ w
\end{array} \right) + \f{t_2}{\sq{2}} \left( \begin{array}{c}
0 \\ 1 \\ w \\ 0
\end{array} \right) \Big] e^{(ik_R-k_I) y} \nn \\
&+& \Big[ \f{t_3}{\sq{2}} \left( \begin{array}{c}
1 \\ 0 \\ 0 \\ -w^*
\end{array} \right) + \f{t_4}{\sq{2}} \left( \begin{array}{c}
0 \\ 1 \\ -w^* \\ 0
\end{array} \right) \Big] e^{(-ik_R-k_I) y}, \nn \\
&& {\rm where} ~~w ~=~ \f{\De_0 k_{ySC}/k_F}{E+\hbar^2\vec k^2/(2m) -\mu_{SC}}.
\label{psi-3-py-dz} \eea

\subsection{Conductance}
From the different reflection and transmission amplitudes calculated
we can compute the four probabilities
$R_N = |r_N|^2$, $R_A = |r_A|^2$, $T_N = |t_N|^2$, and $T_A = |t_A|^2$
as functions of $\th_1$ and $E$. The boundary condition Eqs.~\eqref{bc-s}
and Eqs.~\eqref{bc-p} imposed conserves both probability and charge
currents at the junction. The conservation of the
probability current at the junction implies that
\bea v_1 \sin \th_1 &=& v_1 (\sin \th_1 R_N + \sin \th_{1h} R_A) \nn \\
&&+ v_2 (\sin \th_2 T_N + \sin \th_{2h} T_A) \label{cons} \eea
for each value of $\th_1$. From charge current conservation we can
write down the charge current on SC side in terms of the charge currents
on TI-1 and TI-2 as $J_{3,tot}=J_{1,in}-J_{2,out}$.
The incoming charge current along the $\hat y$ direction on the TI-1
side is equal to $J_{1,in} = e v_1 [\sin \th_1 (1 - R_N) + \sin \th_{1h} R_A]$,
while the total outgoing charge current along the $-{\hat z}$ direction on
the TI-2 side is equal to $J_{2,out} = e v_2 [\sin \th_2 T_N -
\sin \th_{2h}T_A]$.
Using the probability conservation~Eq.~\eqref{cons}, this can be written as-
\beq J_{3,tot} ~=~ 2e (v_1 \sin \th_{1h} ~R_A + v_2 \sin \th_{2h} ~T_A)~.
\label{j3tot} \eeq
Essentially, we have written the charge currents on all the three sides in
terms of only the scattering probabilities in TI-1 and TI-2.

The above discussion needs to be modified if there is an evanescent Andreev
mode on TI-1. When $k_{yh}$ becomes purely imaginary, the term proportional
to $r_A$ in Eq.~\eqref{psi-1} becomes a decaying wave (rather than a plane
wave) and therefore does not contribute to the probability and charge currents
along the $\hat y$ direction. Hence Eqs.~\eqref{cons} and \eqref{j3tot} will
not contain the term proportional to $R_A$. The same argument can be repeated
if there is an evanescent Andreev mode on TI-2; then Eqs.~\eqref{cons} and
\eqref{j3tot} will not contain the term proportional to $T_A$.

To obtain expressions for the various conductances from the above currents,
we assume that a voltage bias $V$ is applied on the TI-1 side maintaining
the TI-2 and SC at the same potential. Namely, we choose the mid-gap energy
on the SC side as $E=0$ and maintain the Fermi energy on TI-2 at zero and
the Fermi energy on TI-1 at $eV$. The differential conductance $G_i=dI_i/dV$
is then the derivative of the current measured on side $i$ ($i=1,2,3$) with
respect to $V$. Integrating the various currents over the angle of incidence
$\th_1$ leads to the following expressions for the differential conductances
at zero temperature,
\bea G_1(E) &=& \frac{G_0}{2} ~\ga_E ~\int_0^\pi d\th_1 ~
[\sin \th_1 (1 - R_N) + \sin \th_{1h} R_A], \nn \\
G_2(E) &=& \frac{G_0}{2} ~\ga_E ~\int_0^\pi d\th_1 ~
\f{v_2}{v_1} [\sin \th_2 T_N - \sin \th_{2h}T_A], \nn \\
G_3(E) &=& G_0 ~\ga_E ~\int_0^\pi d\th_1
[ \sin \th_{1h} R_A + \f{v_2}{v_1} \sin \th_{2h} T_A]~, \nn \\
{\rm where}&& G_0 = \f{2e^2}{h}\f{W\mu_{TI}}{hv_1} ~~{\rm ~and~}~~
\ga_E=1+\f{E}{\mu_{TI}}. \label{g123} \eea
Here, we have written
$G_0$ such that it has the units of conductance and we have chosen
$\mu_{TI}>\De_0$. The factor of $\mu_{TI}+E$ in the product
$G_0 \ga_E$ comes from the linear density of states of the incoming
electrons on the TI-1 side. Finally, for the case $v_2 = v_1$ and
$\mu_{TI} \gg E$ (when we find that $\th_2$, $\th_{1h}$ and
$\th_{2h}$ are all equal to $\th_1$), we find that the conductances
must lie within certain bounds: $0 \le G_1 \le 2G_0$, $-G_0 \le G_2
\le G_0$ and $0 \le G_3 \le 2G_0$.

\section{Conductance Results}~\label{sec-res}
In this section, we compute the sub-gap charge conductance
numerically for a chosen set of parameters for $s$- and $p$-wave
superconductors as worked out in Sec.\ \ref{sec-swave} and
\ref{sec-pwave}. The plots of the sub-gap conductance $G_3$ as a
function of the bias voltage $E$ are shown in Fig.~\ref{fig_GV_s}
and Fig.~\ref{fig_GV_py_dz} for the choice $\chi_1=\chi_2=0$. We
find that in accordance with conventional NM-SC junctions, $G_3$
shows peaks at the gap-edge for the $s$-wave SC and at mid-gap for
the $p$-wave SC. These peaks get sharper with increasing barrier
strengths and for large $\chi_3$, $G_3$ reflects the density of
states of the superconductor. Note that the sharp mid-gap peak for
the $p$-wave case indicates the localized mid-gap edge states. However,
in contrast to conventional NM-SC junctions, the peak height does
not reach a value of $2G_0$ for large $\chi_3$ if $\chi_1 = \chi_2 = 0$.
This behavior is
reminiscent of the graphene-SC junctions and is a reflection of the
Dirac-like properties of the electrons on the TI side \cite{grap1}.
To elucidate this property further, we are going to turn on $\chi_1$
and $\chi_2$ and study the behavior of the sub-gap conductance as
their function. Before resorting to such a study, we plot
$G_{1,2,3}$ for sub-gap voltages and for $\chi_1=\chi_2=0$ in
Figs.~\ref{fig_G123}~(a-b). As discussed earlier, these satisfy the
relation $G_1 = G_2 + G_3$.

\begin{figure}[htb]
\epsfig{figure=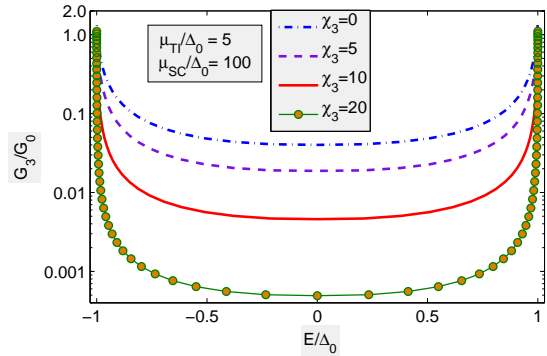,width=8.0cm}
\caption{The sub-gap conductance of an $s$-wave SC in units of $G_0$, for 
$\chi_1=\chi_2=0$. The $G_3$-axis is shown on a logarithmic scale to make the 
distinction between different lines clearer.} \label{fig_GV_s} \end{figure}

\begin{figure}[htb]
\epsfig{figure=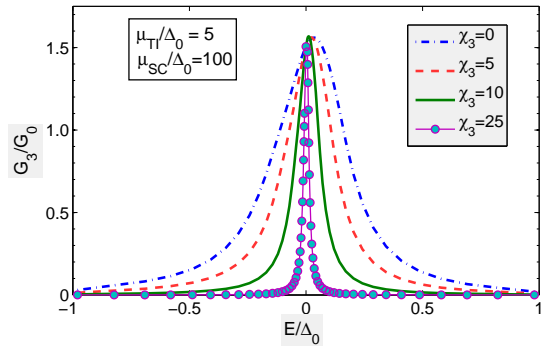,width=8.0cm}
\caption{The sub-gap conductance of a $p_y$-wave SC in units of $G_0$, for 
$\vec d=\hat z$ and $\chi_1=\chi_2=0$.} \label{fig_GV_py_dz} \end{figure}

\begin{figure}[htb]
\subfigure[~$s$-wave SC]{
\epsfig{figure=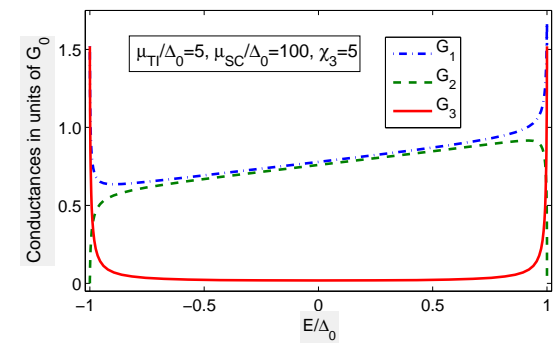,width=8.0cm}}
\subfigure[~$p_y$-wave SC with $\vec d=\hat z$]{
\epsfig{figure=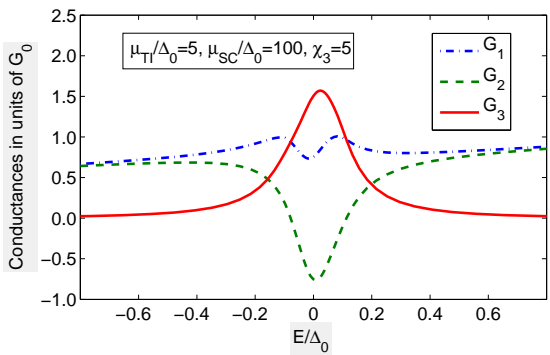,width=8.0cm}}
\caption{The differential conductances $G_3$ into the SC, $G_2$ into the TI-2
and $G_1$ back to the TI-1 in units of $G_0$, for (a) an $s$-wave SC, and (b)
a $p_y$-wave SC with $\vec d=\hat z$.} \label{fig_G123} \end{figure}

\section*{Edge States}
To understand the variation of the sub-gap tunneling conductance
$G_3$ with the barrier potentials $\chi_1$ and $\chi_2$, we first
discuss the localized edge states of the superconductor near the
line junction. We note that a SC with a boundary along $y=0$,
depending on its pairing symmetry, may exhibit sub-gap localized edge
states at the boundary. These states appear at the edges (middle) of
the superconducting gap for $s$- ($p$-) wave pairing symmetry
\cite{hu1}. To understand this, let us first consider a system
consisting of only a SC with a hard wall at $y=0$, and study the
bound states which can occur in that case; later, we shall discuss
the effect that such bound states have on the conductance when the
SC is coupled to the two TIs.

The wave function of a bound state will have a factor $e^{i(k_x x +
k_y y - Et/\hbar)}$, where $k_x$ and $E$ are real and $Im (k_y)
> 0$. Given some real values of $E$ and $k_x$, dispersion relation
will generally have four solutions for $k_y$ of the form $k$, $k^*$,
$-k$ and $-k^*$. Of these we choose the two solutions whose
imaginary part is positive. Let us call those solutions $k_{\pm} =
\pm k_R + i k_I$, where $k_R$ and $k_I$ are positive and denote the
corresponding four-component eigenstates of the Hamiltonian by
$\Psi_+$ and $\Psi_-$ for a particular component of the spin.
If these eigenstates are {\it identical},
then we can take their superposition with opposite signs to obtain a
wave function which is proportional to $\Psi_+ \sin (k_R y) e^{i(
k_x x - Et/\hbar) - k_I y}$; this vanishes at $y=0$ thereby
satisfying the hard wall boundary condition and giving us a bound
state. In general, for a given value of $E$, there will be a
particular value of $k_x$ for which there are two eigenstates which
are identical and can therefore be superposed to give a bound state;
hence the bound states will have a dispersion in which $E$ is a
function of $k_x$. Further, there will be two such states for a
given $E$ and $k_x$ corresponding to the spin degree of freedom.

Now we discuss the effect that the existence of bound states on the
SC-side may have an effect on the conductance when the SC is coupled
to two TIs. In general, an electron incident on TI-1 penetrates into
the SC up to some distance ($\propto 1/k_R$) and the amplitude
(hence the transmission probability) for such a penetration can be
suppressed by increasing the barrier strength $\chi_3$. However, if
an electron is incident with an energy which is exactly equal to the
bound state energy on the SC side, there will be a resonance in the
transmission to the SC side and the barrier $\chi_3$ does not affect
the transmission. A flat dispersion for the bound state spectrum
therefore means that the conductance should be enhanced when the
bias matches exactly the bound state energy on SC side. We shall
soon see that the bound state spectrum is flat for certain cases.
For larger barrier strength $\chi_3$, the peak is more visible since
a larger barrier will suppress transmission at energies other than
the bound state energy. We find such peaks in our numerical
calculations for (i) an $s$-wave SC at $E=\pm\De$ (see
Fig.~\ref{fig_GV_s}) and (ii) a $p$-wave SC at $E=0$ (see
Fig.~\ref{fig_GV_py_dz}) as expected \cite{hu1}.

\noi {\bf $s$-wave SC}: The wave function $\Psi_3$ for an $s$-wave
SC is given by Eq.~\eqref{psi-3s}. We find that the wave~function
$\Psi_3$ consistent with the hard wall boundary exists only when
$E=\pm\De$. The dispersion for the bound states is flat $E=\pm\De$
for any $k_x$. These bound states are responsible for the sub-gap
conductance peaks at $E=\pm \De$ in Fig.~\ref{fig_GV_s}. At $E=\De$,
$\eta=0$ and hence $\Psi_3$ and $\Do_y\Psi_3$ can be written as a
linear combination of the spinors $[1~0~0~1]^T$ and
$[0~~1~-1~~0]^T$. So if we multiply Eqs.~\eqref{bc-s} from the
left by the orthogonal row vectors $[1~~0~~0~-1]$ and $[0~1~1~0]$,
the boundary condition reduces to four equations with four unknowns
$t_N$, $t_A$, $r_N$ and $r_A$ which can easily be solved. (The details of
a similar calculation for a $p$-wave SC are presented in the Appendix). The
important point to note here is that the barrier strength $\chi_3$
drops out of the problem. Let us choose $\chi_1=\chi_2=0$ for
simplicity. We find that $t_N=t_A=0$ and \beq |r_A|^2 =
\f{1-\cos{\th_1}}{1+\cos{(\th_1+\th_{1h})}} \label{rA-s}. \eeq A
similar calculation yields the same expression for the scattering
probabilities for the case $E=-\De$. However the difference to note
is that $\th_{1h}$ is different at energies $E=\pm\De$ since
$\cos{\th_{1h}}=\cos{\th_1} ~(\mu_{TI}+E)/(\mu_{TI}-E)$.
Numerically, we find that the conductance $G_3$ approximately obeys
$G_3(\De)=G_3(-\De)$ in the limit $\mu_{SC}\gg\De$.

\noi {\bf $p$-wave SC}: Let us study a $p$-wave SC with the choice
$f(\vec k)=k_y/k_F$ and $\vec d=\hat z$. For this case, the wave
function $\Psi_3$ given by Eq.~\eqref{psi-3-py-dz} is consistent
with the hard wall boundary only when $w=-w^*$ which simplifies to
\beq \Big[E-\f{\hbar^2(k_x^2+k_R^2)}{2m} +\mu_{SC}\Big]k_R =0
\label{bound-p}\eeq We find that the bound state dispersion is flat:
$E=0$ for all $k_x$. [However when $f(\vec k)\propto k_y+i\al k_x$,
with a non-zero $\al$, the bound states have a dispersion owing to TRS
breaking]. From the calculation detailed in the Appendix, we see
that for $E=0$,
$r_A=t_N=0$,
\beq |t_A|^2 = \f{1-\cos{2\th_1}}{1+\sin^2{\th_1} +\cos^2{\th_1} ~
\sin{[2(\chi_1-\chi_2)]}}, \label{tA-p-dz} \eeq
and $|r_N|^2 = 1 - |t_A|^2$. Remarkably, $|t_A|^2$ and hence the mid-gap
conductance $G_3(0)$ depend only on the difference $\chi_1-\chi_2$.
[This is in contrast to the conductance of a
junction of just two topological insulators (i.e., our set-up
without the SC side) where the total barrier strength seen by the
electrons is $\chi_1+\chi_2$; then the conductance will only depend
on $\chi_1+\chi_2$]. The important result that $|t_A|^2$ depends only
on the combination $\chi_1-\chi_2$ (and hence $G_3$ also,
since $r_A = 0$) at zero bias can be understood qualitatively
as follows. An electron incident on the junction from the TI-1 side
sees a barrier strength of $\chi_1$ on that side. It then transmits to
the TI-2 side with an amplitude $t_A$ as a hole;
such a hole sees a barrier strength of $-\chi_2$ on that side, since
the potentials seen by electrons and holes have opposite signs. Hence
the total barrier strength seen is given by $\chi_1 - \chi_2$; hence
this is the parameter which appears in Eq.~\eqref{tA-p-dz}.

The numerical results shown in Fig.~\ref{fig_G_vs_ch_py_dz}
highlight these features. To illustrate the
behavior of conductance as a function of different barrier
strengths, we choose $\chi_1=\chi_2=\chi_3=\chi$.
At non-zero bias, the conductance has periodic oscillations with a
decaying envelope as shown in Fig.~\ref{fig_G_vs_ch_py_dz}~(a).
The periodic oscillations are due to the barriers $\chi_1$ and
$\chi_2$ on the TI surfaces while the decaying envelope is due to
the barrier $\chi_3$ on the SC side. At zero bias (i.e., $E=0$),
the barrier on the SC side no more plays a role in determining
the conductance as we saw analytically earlier. However, the barriers
$\chi_1$ and $\chi_2$ on the TI side together determine the
value of the ZBP. The numerical results in
Fig.~\ref{fig_G_vs_ch_py_dz}~(a) for $\chi_2=\chi_1$ and (b) for
$\chi_2 = - \chi_1$ show that the value of $G_3$ at $E=0$
is a function of $\chi_1-\chi_2$ only.

\begin{figure}[htb]
\subfigure[~~$\chi_1=\chi_2$,~~$\vec d=\hat z$]{
\epsfig{figure=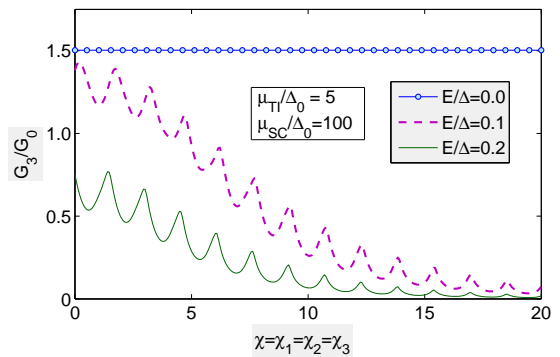,width=8.0cm}}
\subfigure[~~$\chi_1=-\chi_2$, $E=0$]{
\epsfig{figure=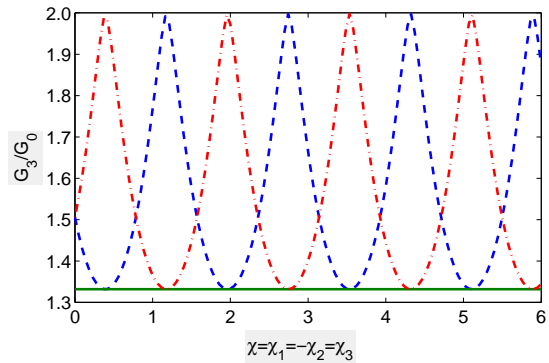,width=8.0cm}}
\caption{(a)~Conductance of a $p_y$-wave SC with 
$\vec d=\hat z$ at different energies in the SC gap. (b)~Conductance of a
$p_y$-wave SC at $E=0$ for different spin pairings: $\vec d=
\hat z$ (blue dashed line), $\vec d=\hat x$ (green solid line)
and $\vec d=\hat y$ (red dot-dashed line). Note that $G_3/G_0$ reaches
2 at $\chi = n\pi/2 + 3\pi/8$ for $\vec d=\hat z$ and at $\chi = n\pi/2
+ \pi/8$ for $\vec d=\hat y$, as can be shown from Eqs.~\eqref{tA-p-dz}
and \eqref{tA-p-dy}.} \label{fig_G_vs_ch_py_dz} \end{figure}

For a $p$-wave SC, when $\vec d=\hat x$ or $\vec d=\hat y$, we find
that the dependence of the conductance on the bias is different from
that for $\vec d=\hat z$ shown in Fig.~\ref{fig_GV_py_dz}. For the
case $\vec d=\hat y$, we again find that $r_A=t_N=0$ while \beq
|t_A|^2 = \f{1-\cos{2\th_1}}{1+\sin^2{\th_1} -
\cos^2{\th_1}~\sin{[2(\chi_1-\chi_2)]}}. \label{tA-p-dy}\eeq (The
qualitative reason for the dependence on $\chi_1 - \chi_2$ is
similar to the one given above for $\vec d = \hat z$). For the case
$\vec d=\hat x$, we find that $t_A=t_N=0$, while
\beq |r_A|^2 = \sin^2{\th_1} \label{rA-p-dx} \eeq
and $|r_N|^2 = \cos^2 \th_1$ are independent of $\chi_1$ and $\chi_2$. This
independence stems from following reason: an electron incident on
the junction from the TI-1 side sees a barrier strength of $\chi_1$.
When it is Andreev reflected as a hole back to that side, it sees a
barrier strength of $- \chi_1$. Hence the total barrier strength
that is seen is zero, regardless of the value of $\chi_1$. Further,
there is no dependence on $\chi_2$ since the electrons does not
transmit to the TI-2 side at all; both $t_A$ and $t_N$ are zero.

These analytical results are confirmed by our numerical
calculations. This is illustrated in
Fig.~\ref{fig_G_vs_ch_py_dz}~(b) where $G_3$ is plotted as a
function of $\chi_1 - \chi_2$ at $E=0$ for the three different forms
of $\vec d$. Note that for $\vec d = \hat y$ and $\hat z$, $G_3$
varies with a period of $\pi/2$ in the variable $\chi = (\chi_1 -
\chi_2)/2$ in Fig.~\ref{fig_G_vs_ch_py_dz}~(b). This is in agreement
with Eqs.~\eqref{tA-p-dz} and \eqref{tA-p-dy} which are periodic in
$\chi_1 - \chi_2 = \pi$. We note that our results demonstrate that
the barrier dependence of the sub-gap tunneling conductance of
junctions of TI and triplet superconductors can be used to determine
the direction of the direction of ${\vec d}$, or equivalently, the
direction of the spin of the Cooper pair. This property is in
complete contrast to the conventional NM-SC junctions and stems from
the spin-momentum locking of the Dirac electrons which breaks the
rotational symmetry in the spin space necessary to distinguish
between the different orientation of ${\vec d}$.

For a $p$-wave SC, we can in general take $f(\vec k)
\propto k_y+i\al k_x$. The choice $\al=0$ corresponds to
$p_y$-SC and the bound state dispersion is flat as discussed
earlier. When $\al$ is changed slightly from zero, we find that
the bound state dispersion develops a small non-zero slope.
Now, the ZBP which existed for $\al=0$ broadens and the peak
value also decreases; the peak eventually disappears as $\al$ is
increased further.

\section{Spin Currents}~\label{sec-spcurr}
In this section we will study the spin currents on the SC side for
$s$- and $p$-wave cases. For $a=x,y,z$, the $a$ component of the
spin density is given by \beq \rho_a ~=~ \f{\hbar}{2} ~\Psi_3^\dg
\tau^z \ot \si^a \Psi_3 \label{rhoa} \eeq in the four-component
language. The operator $\tau^z$ appears in Eq.~(\ref{rhoa}) because
a hole with spin-up (down) corresponds to a missing electron with
spin-down (up). The spin current $\vec J_a$ corresponding to the
spin density $\rho_a$ can be calculated on the SC side using the
equations of motion Eq.~(\ref{eoms}) and (\ref{eomp}) and the
continuity equation $\Do_t\rho_a + \vec \nabla \cdot {\vec J_a} =0$.
The equation of continuity implies that in a steady state, the
component ${\hat y} \cdot {\vec J_a}$ is independent of the $y$
coordinate and can therefore be evaluated at any convenient point
like $y=0$ or $\infty$. We are interested in only the $\hat y$
component of the $a$-spin currents since only this component
contributes to the spin density that is injected into the SC. The
differential spin conductance $G^s_a$ corresponding to the spin
density $\rho_a$ is obtained by integrating over the spin current
contributions ${\hat y} \cdot {\vec J_a}$ from all the angles of
incidence $0<\th_1<\pi$~: \bea G^s_a &=& G^s_0 ~\ga_E
~\Big[\f{k_F}{\mu_{SC}}\int_0^{\pi} d\th_1~\hat y\cdot \vec J_a
\Big] \label{gs}\eea where $G^s_0 = \f{e}{h} ~\f{W\mu_{TI}}{h v_1^2}
~\f{\mu_{SC}}{k_F}$. Here $G^s_0$ has been defined in such a way
that the term in the brackets in the above equation becomes
dimensionless.

In a normal metal, i.e., setting $\Delta_0=0$ in Eq.~(\ref{eoms}), we find that
\beq \vec J_a ~=~ \f{\hbar^2}{2m} ~Im [\Psi_3^{\dg}\si^a \vec \nabla \Psi_3].
\label{ja1} \eeq
In a SC, the spin current $\vec J_a$ is a sum of two parts: the part
$\vec J_{a,N}$ which is independent of $\Delta_0$ given by Eq.~(\ref{ja1}),
and the part $\vec J_{a,\De}$ proportional to $\Delta_0$. $\vec J_{a,N}$
is the expression for the $a$-spin current in a normal metal carried by
either electrons or holes; in SC it gets contribution from the BdG
quasiparticles.
For an $s$-wave SC, we find that the $x$- and $z$- spin currents (i.e.,
$\vec J_a$ for $a=x,z$) are entirely given by Eq.~(\ref{ja1}) and do not
contain the SC pair potential $\De_0$ ($J_{x/z,\De}=0$). Since the
wave function $\Psi_3$ decays exponentially as $y\to \infty$, the spin
current going into the SC is zero at $y \to \infty$ (and therefore at
any value of $y$) if $a=x$ or $z$.
But the $y$-spin current in addition to the expression in Eq.~(\ref{ja1}),
also contains a part proportional to $\Delta_0$ given by-
\beq \hat y \cdot \vec J_{y,\De} = - \Delta_0
~\int_0^y dy' ~ Re [\Psi_3^\dg (x,y')\tau^x \Psi_3 (x,y')].
\label{jas} \eeq
The derivation of this is similar to that of Eq.~(\ref{jpair}). Hence
the total spin current is given by the sum of Eqs.~(\ref{ja1}) and
(\ref{jas}), and this need not vanish. It is easiest to evaluate this at
$y=0$ where Eq.~(\ref{jas}) vanishes. For $a=y$, after tedious calculation,
we find that Eq.~(\ref{ja1}) simplifies to
\beq \hat y \cdot \vec J_y = \f{\hbar^2 k_R}{2m}(1+|w|^2)~
Im[t_1^*t_2-t_3^*t_4]. \label{jys}\eeq
The above quantity is non-zero for the scattering of an electron incident
at energy $E$, at an angle $\th_1$ on TI-1. But the $\hat y \cdot \vec J_y$
currents at incident angles $\th_1$ and $\pi-\th_1$ add up to zero, thus
contributing nothing to the differential spin conductance $G^s_y$. To
summarize, all the spin conductances $G^s_a$ are zero for the $s$-wave case.

For a $p$-wave SC, the $a$-spin current is given by
\bea \vec \nabla \cdot \vec J_{a,\De} = \De_0~ &Re\Big[& \vec a
\cdot [ \psi^\dg \si^a (\vec d \cdot \vec\si) i\si^y \vec\nabla\phi \nn \\
&&- \vec\nabla\psi^{\dg} (\vec d \cdot \vec\si) i\si^y \si^a \phi] \Big].
\label{j3pair}\eea
We will now consider all the nine different possibilities corresponding
to $\vec d = \hat x, \hat y, \hat z$ and $a = x,y,z$. Table~\ref{tab-js}
summarizes the results for different spin currents, while Table~\ref{tab-gs}
summarizes the spin conductance results (obtained by integrating out
spin currents over all incident angles $\th_1$) for different cases.
For the four cases corresponding to $(\vec d, a) = (\hat x, x)$,
$(\hat x, y)$, $(\hat z, y)$ and $(\hat z, z)$, we find that
$\vec \nabla \cdot \vec J_{a,\De}$ is a total derivative, which
implies that $\vec J_{a,\De}$ has a local expression in terms of
$\Psi_3$ (analogous to the last terms in Eq.~(\ref{jpp})). Then
one can evaluate both $\vec J_{a,\De}$ and
Eq.~(\ref{ja1}) at $y=\infty$; they vanish there because $\Psi_3$ goes
to zero exponentially. Hence no spin current enters the SC in these four
cases. In the remaining five cases, we find that $\vec \nabla \cdot
\vec J_{a,\De}$ is not a total derivative; hence $\vec J_{a,\De}$ has
an integral expression (analogous to Eq.~(\ref{jas})). Since this
vanishes at $y=0$, one can find the total spin current by evaluating
only Eq.~(\ref{ja1}) at $y=0$.

\begin{table}[htb]
\begin{center}
\begin{tabular}{|c|c|c|c|c|}
\hline
~ & $s$-wave& $p$-wave & $p$-wave & $p$-wave \\
~ &~& $\vec d=\hat x$ & $\vec d=\hat y$ & $\vec d=\hat z$ \\
\hline
$J_x$ & 0 & 0 & Eq.~\eqref{jxdy} & Eq.~\eqref{jxdy} \\
\hline
$J_y$ & Eq.~\eqref{jys} & 0 & Eq.~\eqref{jydy} & 0 \\
\hline
$J_z$ & 0 & Eq.~\eqref{jzdx} & Eq.~\eqref{jzdx} & 0 \\
\hline
\end{tabular}
\end{center}
\caption{Expressions for spin currents for $s$-wave and $p$-wave
pairing for a particular angle of incidence.} \label{tab-js} \end{table}

\begin{table}[htb]
\begin{center}
\begin{tabular}{|c|c|c|c|c|}
\hline
~ & $s$-wave& $p$-wave & $p$-wave & $p$-wave \\
~ & ~& $\vec d=\hat x$ & $\vec d=\hat y$ & $\vec d=\hat z$ \\
\hline
$G^s_x$ & 0 & 0 & non-zero & non-zero \\
\hline
$G^s_y$ & 0 & 0 & 0 & 0 \\
\hline
$G^s_z$ & 0 & 0 & 0 & 0 \\
\hline
\end{tabular}
\end{center}
\caption{Spin conductances $G^s_a$ for $s$-wave and $p_y$-wave SC.}
\label{tab-gs} \end{table}

\noi {\bf $\vec d = \hat x$~}: As argued earlier, no $a$-spin
current corresponding to $a=x,y$ enters the SC. $J_z$ on the other
hand has an expression given by \beq \hat y \cdot \vec J_z =
\f{\hbar^2 k_R}{4m} (1+|w|^2)~[ |t_1|^2-|t_2|^2-|t_3|^2+|t_4|^2]~.
\label{jzdx} \eeq
This is the expression for $z$-spin current for
electron incident at a given energy $E$ and angle $\th_1$ on TI-1.
The $z$-spin conductance on SC is obtained by summing up the
currents at all possible incident angles $0< \th_1 <\pi$. The $z$-spin
conductance $G^s_z$ turns out to be zero due to the exact cancellation
of the $\hat y \cdot \vec J_z$ at angles $\th_1$ and $\pi - \th_1$.

\noi {\bf $\vec d = \hat y$~}: All the three spin currents could have
a non-zero value in this case~(see Table~\ref{tab-js}). The expression
for $z$-spin current in this case is given by Eq.~\eqref{jzdx} and the
total $z$-spin current at any given energy $E$ in the gap goes to zero
by the same logic that follows Eq.~\eqref{jzdx}. The expressions for
the $x/y$-spin currents are
\bea \hat y \cdot \vec J_{x} = \f{\hbar^2k_R}{2m}(1+|w|^2)~Re[t_1^*t_2
-t_3^*t_4],\label{jxdy} \\
\hat y \cdot \vec J_{y} = \f{\hbar^2k_R}{2m}(1+|w|^2)~Im[t_1^*t_2
-t_3^*t_4].\label{jydy}\eea
Integrating the spin current over $\th_1$, we find that the $y$-spin
conductance is zero while the $x$-spin conductance is non-zero and has
features similar to the $\vec d =\hat z$ case as discussed below.

\noi {\bf $\vec d = \hat z$~}: In this case as argued earlier,
$y/z$-spin currents are zero. The expression for $x$-spin current is
given by Eq.~\eqref{jxdy}. The spin conductance which is integrated
over $x$-spin current is non-zero and has some feature as shown in
Fig.~\ref{fig_GsV_py_dz} for a typical case.

If we compare Tables \ref{tab-js}~and~\ref{tab-gs}, the non-zero
$y$($z$)-spin current for an electron incident at angle $\th_1$ in
Table~\ref{tab-js} adds up with the $y$($z$)-spin current for an
electron incident at angle $\pi-\th_1$ to give zero in the total
$y$($z$)-spin conductance. This can be seen from the symmetry of the
equations of motion and the boundary conditions under the
transformation $x\to -x$. For both TI-1 and TI-2, the equations of
motion (Eq.~\eqref{eomti}) remain invariant under the
transformations: $\Psi_i\to\si^x\Psi_i$ and
$\Psi_i\to\tau^z\si^x\Psi_i$ accompanied by $x \to -x$. Both the
boundary conditions Eqs.~\eqref{bc-s} and Eqs.~\eqref{bc-p} are
invariant under these two transformations. However, for $s$-wave SC
and the $p$-wave SC with $\vec d=\hat x$ the equations of motion
(Eq.~\eqref{eoms} and Eq.~\eqref{eomp}) are invariant under the
transformation $\Psi_3\to\tau^z\si^x\Psi_3$. And for $p$-wave SC
with $\vec d=\hat y,~\hat z$ the equations of motion
(Eq.~\eqref{eomp}) are invariant under the transformation
$\Psi_3\to\si^x\Psi_3$. Now, we can easily see from Eq.~\eqref{ja1}
that under both the transformations~($\Psi_i\to\si^x\Psi_i$ and
$\Psi_i\to\tau^z\si^x\Psi_i$), $\hat y \cdot \vec J_x$ is invariant
while $\hat y \cdot \vec J_y$ and $\hat y \cdot \vec J_z$ change
sign. Since $x\to-x$ is same as $\th_1\to \pi-\th_1$, this explains
the exact cancellation of $y$- and $z$-spin currents for $\th_1$ and
$\pi-\th_1$.

We can relate the spin currents which are non-zero
in Table~\ref{tab-gs} to the physical quantities
on the other side of the junction, i.e., on the TI-1 and TI-2 sides.
Using the boundary conditions in Eqs.~\eqref{bc-p}, we can rewrite the
$x$-spin current at the junction as
\beq \hat y \cdot \vec J^{SC}_x ~=~ \f{\hbar}{2}[v_1\Psi_1^{\dg}\tau^z\Psi_1
- v_2\Psi_2^{\dg}\tau^z\Psi_2 ] \label{js_cons}\eeq
for a $p_y$-wave SC with $\vec d= \hat y$ or $\hat z$.
We thus see that the $x$-spin current on the SC side of the junction
is linearly related to the steady state charge densities on the TI-1
and TI-2 sides of the junction evaluated at the junction.

\subsection*{Spin Conductance}
The $x$-spin conductance shows an unusual satellite~peak~(SP) in
addition to the ZBP for the cases $\vec d=\hat y$ and $\vec d=\hat z$
when $f(\vec k)=k_y/k_F$~(see Fig.~\ref{fig_GsV_py_dz}).
This SP merges with the ZBP for $\chi_3 \gtrsim 20$.
The SP is observable only if there is an appreciable penetration of
a plane wave state with non-zero energy from the TI-1 into the SC,
and this can occur only if $\chi_3$ is not too large. The location
and height of the SP change with $\chi_1$ and $\chi_2$ in a periodic
way. To highlight these features, we choose $\chi_1=\chi_2=\chi_3
\equiv\chi$ and show the spin conductance $G^s_x$ as a function
of the bias and the barrier strength $\chi$ in Fig.~\ref{fig_Gs_E_ch}
as a contour plot.

We find, using Eq.~\eqref{js_cons}, that the $x$-spin current in
the superconductor can be written in terms of the charge densities
in TI-1 and TI-2. In the case $v_1=v_2$, we find that the spin
current is proportional to
$\Psi_1^{\dg}\tau^z\Psi_1-\Psi_2^{\dg}\tau^z\Psi_2$ which is given by
\bea && \Psi_1^{\dg}\tau^z\Psi_1-\Psi_2^{\dg}\tau^z\Psi_2 \nn \\
&& = 1+R_N+R_{N,Ph}-R_A\Ga_{1A}-T_N+T_A\Ga_{2A}, \label{gs12}\eea
where
\bea R_{N,Ph} &=& Re (r_N+r_N e^{-i2\th_1}), \nn \\
\Ga_{1A} &=& \nu_E^2\cos^2{\th_1}-
\nu_E\cos{\th_1}\sq{\nu_E^2\cos^2{\th_1}-1}, \nn \\
\Ga_{2A} &=& |\nu_E\cos{\th_1}|, \label{gss12} \eea
in the range of
$(E,\th_1)$ for which there exist evanescent Andreev modes on TI-1
and TI-2, i.e., when $\nu_E\cos{\th_1}>1$. When the Andreev states
are plane wave states on TI-1 and TI-2, $\Ga_{1A}$ and $\Ga_{2A}$
both are equal to $1$ in Eq.~\eqref{gs12}.

\begin{figure}[htb]
\epsfig{figure=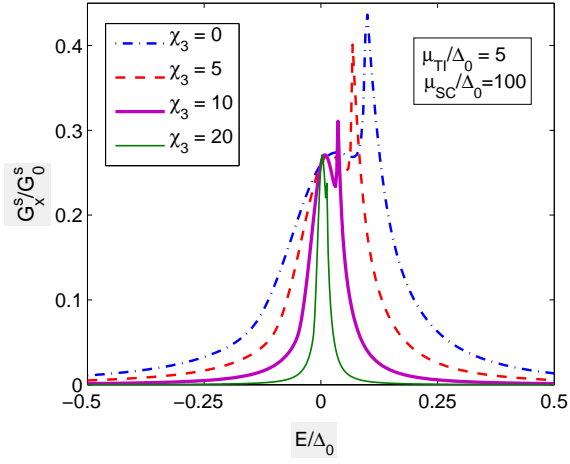,width=8.0cm}
\caption{Spin conductance for a $p_y$-wave SC with $\vec d=\hat z$ and 
$\chi_1=\chi_2=0$. There is a satellite peak at a positive value of the bias 
which merges with the ZBP in the limit of large $\chi_3$.}
\label{fig_GsV_py_dz} \end{figure}

\begin{figure}[htb]
\epsfig{figure=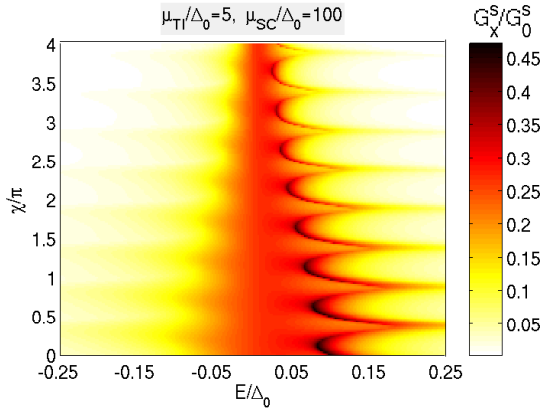,width=8.0cm}
\caption{Spin conductance for a $p_y$-wave SC with $\vec d=\hat z$ and 
$\chi_1=\chi_2=\chi_3=\chi$. Both the location and 
the height of the SP oscillate and decay with increasing $\chi$.}
\label{fig_Gs_E_ch} \end{figure}

Using the identity in Eq.~\eqref{gs12}, we find the contributions to the spin
conductance $G_x^s$ from the different scattering amplitudes on the TI-1
and TI-2 sides. It is interesting to note that in contrast to the
expressions in Eq.~\eqref{g123} for the charge conductances where
only the scattering probabilities in the TI-1 and TI-2 appear, the
spin conductance on SC side depends both on the phase and the
amplitude of the reflection amplitude $r_N$. We show in
Fig.~\ref{fig_GsV_case1} the contributions to the spin conductance
from the different terms in Eq.~\eqref{gs12}. From the figure, it is
evident that the Andreev scattering term $(T_A \Ga_{2A} - R_A
\Ga_{1A})$ and the term $R_{N,Ph}$ (which is proportional to $r_N$)
contribute the most to the SP. Further, to understand the origin of
the SP, we look at the spin currents due to incident electrons with
different energies and angles of incidence $\th_1$. In
Fig.~\ref{fig_Gs_Eth}, we show spin conductances at different $E$
and $\th_1$ as a contour plot. This contour plot shows that the ZBP
and the SP in the spin conductance get contributions from different
angles $\th_1$. In contrast to the ZBP in the spin conductance which
gets contributions mainly around normal incidence ($\th_1\sim
\pi/2$), the SP gets major contributions from electrons incident at
glancing angles ($\th_1\sim 0$ and $\th_1\sim \pi$). This is
highlighted in the contour~plot by ellipses. The green~(dot-dash)
lines, $\th_1 = \cos^{-1}( \pm 1/\nu_E)$, separate the evanescent
Andreev modes on TI-1 and TI-2 from the Andreev plane wave
states. The largest contribution to the SP comes from the range of
$\th_1$ for which there are evanescent Andreev modes; this is shown
by the pink~(dark) region in the contour plot. In the same region,
we find that the phase of $r_N$ changes rapidly from $-\pi$ to $\pi$
(going through $0$) as $E$ is varied at a fixed value of $\th_1$.
Such a feature in the phase of $r_N$ shows up as a peak in the term
$R_{N,Ph}$ in Eq.~\eqref{gs12} when $E$ is varied at a fixed
$\th_1$. Further, the SP in spin conductance merges with the ZBP in
the limit $\mu_{SC}\to\infty$.

\begin{figure}[htb]
\epsfig{figure=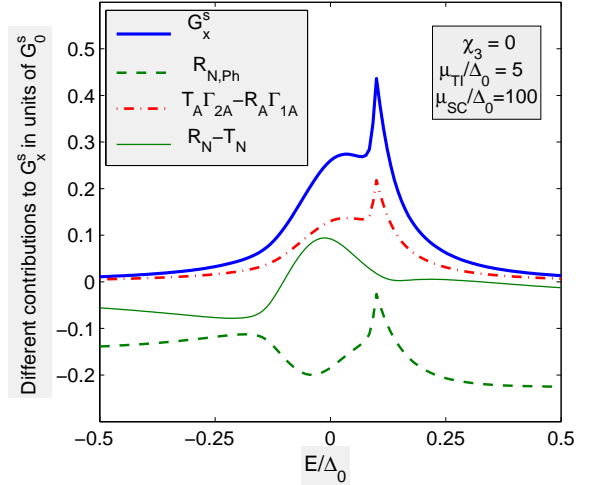,width=8.0cm}
\caption{Spin conductance and the contributions from different terms in
Eq.~\eqref{gs12} for a $p_y$-wave SC with $\vec d=\hat z$ and $\chi_1=\chi_2
=\chi_3=0$. The phase term $R_{N,Ph}$ and the Andreev terms $T_A\Ga_{2A} -
R_A\Ga_{1A}$ contribute to the SP.} \label{fig_GsV_case1} \end{figure}

\begin{figure}[htb]
\epsfig{figure=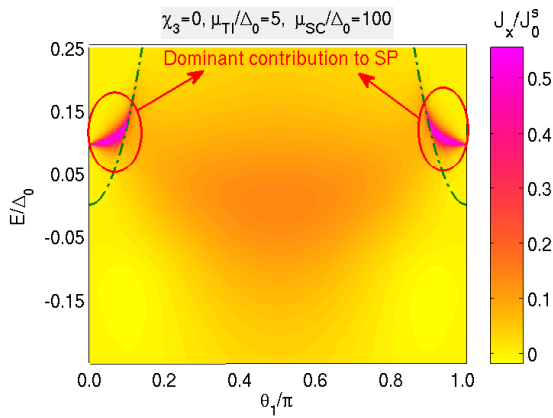,width=8.0cm}
\caption{Contour plot of spin current $\hat y\cdot \vec J_x$ (in units of
$J^s_0=\hbar^2k_F/(2m)$) as a function of the energy and incident angle of 
the electron on the TI-1 side,
for a $p_y$-wave SC with $\vec d=\hat z$ and $\chi_1=\chi_2=\chi_3=0$.
The dominant contribution to the SP comes from the pink~(dark) regions
which are bounded by the green dot-dash lines. See text for details.}
\label{fig_Gs_Eth} \end{figure}

There is a striking asymmetry about $E=0$ in the spin conductance in
Fig.~\ref{fig_GsV_py_dz}; this mainly arises because the evanescent Andreev
modes can exist only for $E>0$ and are absent for $E<0$.
There is also an asymmetry due to a finite difference in
the density of states of the incident electrons at energies $E$ and
$-E$~(see Eq.~\eqref{g123}) and the lack of invariance of
Eq.~\eqref{psi-3-py-dz} under $E\to-E$. A small asymmetry is also visible in
the charge conductance shown in Fig.~\ref{fig_GV_py_dz}.

\section{Summary and Discussion}~\label{sec-disc}
In this paper, we have considered a junction between two TIs and a SC
which can be either $s$-wave or $p$-wave; a $p$-wave SC may have
different spin pairings and we have studied each case. We have formulated
the most general time-reversal invariant boundary condition at the junction;
this consists of three barrier parameters. We have studied
the dependence of the differential charge and spin conductances on the
barrier parameters, the ratio of the TI and SC chemical potentials to the
superconducting pair potential, and the voltage bias.

Our main results can be summarized as follows.

\noi (i) In accordance with conventional NM-SC junctions, we find
that the charge conductance $G_3$ shows peaks at the edges of the SC
gap for $s$-wave SC; for a $p$-wave SC, there is a peak at the zero
bias. However, in contrast to conventional junctions, the heights of
these peaks do not reach the conventional value of $2G_0$, if $\chi_1 =
\chi_2 = 0$; this
feature is a signature of the Dirac nature of the TI quasiparticles.
Further, the height of these peaks is independent of $\chi_3$. The
conductance peaks arise due to the presence of bound states on the
SC~side; as a result, an electron coming in from the TI side with
the same energy as the bound state can resonantly enter the SC, and
then convert into Cooper pairs deep inside the SC. (The phenomenon
of a conductance peak resembles the transmission resonance which
occurs in other systems; see Ref.~\onlinecite{reso} for a few
examples).

\noi (ii) For a $p$-wave SC, we studied the conductance as a
function of barrier strengths. At non-zero energies the conductance
both oscillates and decreases with increasing barrier strength. This
demonstrates both the Dirac and the Schr\"odinger nature of the
electrons in different parts of our set-up in the following sense.
The transmission of Dirac electrons oscillates with the barrier
strengths in the TI-1 and TI-2 (as can be seen by the presence of
$\cos \chi_i$ and $\sin \chi_i$ terms in the boundary
conditions~(\ref{bc-s}) and (\ref{bc-p}), for $i=1,2$), while the
transmission of Schr\"odinger electrons through the barrier in the SC
decays exponentially with the barrier strength $\chi_3$. This indicates that
for $\chi_1 = \chi_2 = 0$ the charge conductance in these junctions does not
reach the maximum value of $2G_0$ which is found in conventional junctions.

\noi (iii) For a $p$-wave SC, the dependence of the conductance on
the barrier parameters (specifically on the difference of the
parameters in the two TIs, $\chi_1 - \chi_2$, at zero bias) varies
with the direction of $\vec d$, i.e., the spin pairing of the Cooper
pairs. This is due to the spin-momentum locking of the electrons in
the TIs and the fact that the geometrical arrangement of the two TIs
completely breaks the rotational symmetry. This is quite different
from what happens in a junction of normal metals with a $p$-wave SC;
since a metal is isotropic in spin space, the conductance into a SC
would not depend on the direction of $\vec d$. Thus these junctions
are expected to provide a test bed for determining the orientation
of $\vec d$ and hence the Copper pair spin direction of 
triplet SC without application of external magnetic fields~\cite{mack}.

\noi (iv) The spin conductance into the SC for different pair
potentials exhibits several unconventional features. We find that
there is a non-zero conductance for the $x$ component of the spin
only for a $p_y$-wave SC in which $\vec d = \hat y$ or
$\hat z$. (The fact that all the other spin conductances are zero
can be shown using a parity symmetry which is present in our system). 
The non-zero spin conductance arises because the TIs break the spin 
rotation symmetry, and it does not occur for a NM-SC 
junction. Most remarkably, the spin conductance shows a satellite
peak away from zero energy. We have provided an explanation of this
satellite peak by relating the spin conductance on the SC side to
the charge densities on the TI sides. The charge density turns out
to depend on the {\it phase} of the normal reflection amplitude
which shows a peak at non-zero bias. The satellite peak in the spin
conductance is a distinctive feature of our system and it would be
interesting to look for this experimentally.

The simplest experimental verification of our theory would be to
measure the maxima of the charge conductance peaks for both $s$- and
$p$-wave SCs. A prediction of our theory is that the values of
these peaks reach $2G_0$ only for some non-zero barrier strengths. It will
also be interesting to measure the barrier potential dependence of $G_3$.
The barrier potentials can be tuned by applying a gate voltage along the
junction in our system. It may be difficult to tune $\chi_1$,
$\chi_2$ and $\chi_3$ separately. Looking at Fig.~\ref{fig_schem},
however, it is apparent that if a gate voltage is applied from above the
junction, it will have a larger effect on $\chi_1$ and $\chi_3$ than on
$\chi_2$ (since it is further away from the TI-2); hence changing such a
gate voltage will change the value of $\chi_1 - \chi_2$. Hence our important
prediction that the zero bias peak depends on $\chi_1 - \chi_2$ for $\vec d
= \hat y$ and $\hat z$ but does not depend at all on any of the $\chi_i$
for $\vec d = \hat x$ is something which can be tested by changing the gate
voltage. A consequence of this would be that measurements of $G_3$ can be
used to determine the direction of $\vec d$ for a given $p$-wave SC by
attaching it to a junction of two TIs.

Recently, there has been much excitement about a zero-bias peak~(ZBP)
observed in a number of experiments in semiconducting/superconducting
nanowires~\cite{mourik,deng,rokhinson,das}. This peak is believed to be
due to a Majorana fermion mode, and it has a close connection to the mid-gap
(zero energy) states which are present at the boundary of a $p$-wave
SC~\cite{kseng01,kseng02,kwon}. The last few years have witnessed
intense theoretical activity in the area of Majorana modes at the ends of
one-dimensional systems~\cite{lutchyn1,oreg,brouwer1,degottardi,
potter,fulga,stanescu,tewari,gibertini,lim,tezuka,egger,ganga,lobos,lutchyn2,
fidkowski,cook,pedrocchi,sticlet,jose, klinovaja,alicea}. In our
system, we have studied the edge states which occur at $E=0$ in a
$p$-wave SC with a hard wall. However, although these
states give rise to a zero bias peak in the charge conductance,
they cannot be called Majorana modes for the following reason.
A Majorana mode must have a real wave function so that it remains invariant
under complex conjugation. However, our zero energy states carry
finite momentum $k_x$ and do not have real wave functions.
Under complex conjugation, a state with momentum $k_x$ changes to a state
with momentum $-k_x$ and therefore does not remain invariant.

Finally, we point out that in this work we have studied the charge
and spin conductances of a junction between two TI surfaces and one
SC. It is natural to ask what would happen if there was
a junction of a single TI surface and an SC. A
peculiarity of this problem would be that a single TI has only one
spin degree of freedom for a given value of the energy and momentum,
while a SC (or metal) has two spin degrees of freedom.
Hence boundary conditions such as the ones given in
Eqs.~(\ref{bc-s}-\ref{bc-p}) would generally be inconsistent since
they provide eight equations for six amplitudes (normal and Andreev
reflection in the TI and four amplitudes on the SC side). For our
system with two TIs and a SC, there is no such mismatch.
This enables us to find all the amplitudes for all values of the
system parameters, and the amplitudes always satisfy the
conservation of probability.

\acknowledgments
A.S. thanks CSIR, India for financial support. D.S. thanks DST, India
for support under grant SR/S2/JCB-44/2010.

\section*{Appendix}
Here, we shall present a calculation to show how different scattering
amplitudes become independent of the barrier $\chi_3$ on the SC side
when the SC side hosts bound states. In particular, we choose to discuss
the case of $p_y$-SC with $\vec d=\hat z$, and $v_1 = v_2$ in the TIs.

From the Hamiltonian in Eq.~\eqref{HSC}, it is easy to see that for a $p_y$-SC
with $\vec d=\hat z$, $w$ defined in Eq.~\eqref{psi-3-py-dz} satisfies
the $2 \times 2$ matrix equation
\beq
\Big[\Big(\f{\hbar^2(k_x^2+k_y^2)}{2m}-\mu_{SC}\Big)\tau^z
+ \f{\De_0k_y}{k_F}\tau^x \Big] \left(\begin{array}{c} 1 \\
w \end{array} \right)
= E \left(\begin{array}{c} 1 \\ w \end{array} \right). \eeq
At $E=0$ we can see that the two rows of this equation imply that
$w^2=-1$. Further, $k_y=k_{ySC}$ given by Eq.~\eqref{kySC-py-dz}
can be written as $k_y=k_R+ik_I$ with $k_I>0$.
With this, it is easy to see that $w=-i$. Then, Eq.~\eqref{psi-3-py-dz}
shows that the wave function on the SC side
$\Psi_3(y)$ and its derivative $\Do_y\Psi_3(y)$
both reduce to a linear combination of only the two spinors
$\Psi_{p\ua}=[1~~0~~0~-i]^T$ and $\Psi_{p\da}= [0~~1~-i~~0]^T$. Also, the
third term on the left hand side (LHS) of the
second boundary condition in Eqs.~\eqref{bc-p} will be a linear
combination of the spinors $\Psi_{p\ua}$ and $\Psi_{p\da}$ for a $p_y$-SC
with $\vec d=\hat z$. Hence the LHS of both the equations in the boundary
condition in Eqs.~\eqref{bc-p} are linear combinations of the
spinors $\Psi_{p\ua}$ and $\Psi_{p\da}$. Since both $\Psi_{p\ua}^T$
and $\Psi_{p\da}^T$ are orthogonal to $\Psi_{p\ua}$ and $\Psi_{p\da}$,
multiplying by $\Psi_{p\ua}^T$ and $\Psi_{p\da}^T$ from the left in
Eqs.~\eqref{bc-p} makes the LHS equal to zero; hence the equations become
independent of $\chi_3$. We then get the following
four equations containing four unknowns $r_N$, $r_A$, $t_N$ and $t_A$,
\bea \Psi_{p\ua}^T \big[M(\chi_1)\Psi_1 + M(-\chi_2)\Psi_2\big] &=& 0, \nn \\
\Psi_{p\da}^T \big[M(\chi_1)\Psi_1 + M(-\chi_2)\Psi_2\big] &=& 0, \nn \\
\Psi_{p\ua}^T \si^x\tau^z \big[M(\chi_1)\Psi_1 - M(-\chi_2)\Psi_2\big] &=&
0, \nn \\
\Psi_{p\da}^T \si^x\tau^z \big[M(\chi_1)\Psi_1 - M(-\chi_2)\Psi_2\big] &=&
0, \eea
where $\Psi_1$ and $\Psi_2$ are evaluated at $y=0$.
When $\Psi_1$ and $\Psi_2$ from Eqs.~\eqref{psi-1}~and~\eqref{psi-2} are
substituted in the above equations, we get four equations in four unknowns
$r_N$, $r_A$, $t_N$ and $t_A$ which can be cast in the matrix form:
${\mathds M} \cdot X = A$ where 
\bea {\mathds M} &=& \left[
\scalemath{0.8}{\begin{array}{c c c c}
g(\chi_1) & - w \ga_{\th1}^* g^*(\chi_1) &  h(\chi_2) & w h^*(\chi_2) \\
-\ga_{\th1}^* g^*(-\chi_1) & w g(-\chi_1) &  -ih(\f{\pi}{2}+\chi_2) &
-iwh(\f{\pi}{2}-\chi_2) \\
-\ga_{\th1}^* g^*(-\chi_1) & -w g(-\chi_1) &  ih(\f{\pi}{2}+\chi_2) &
-iwh(\f{\pi}{2}-\chi_2) \\
g(\chi_1) &  w \ga_{\th1}^* g^*(\chi_1) &  -h(\chi_2) & w h^*(\chi_2) \\
\end{array}} \right], \nn \\
X &=& \scalemath{1.0}{[r_N ~~~r_A ~~~t_N ~~~t_A]^T}, \nn \\
A &=& [\scalemath{0.8}{ -g^*(\chi_1) ~~-ig^*(\f{\pi}{2}+\chi_1) ~~-ig^*(
\f{\pi}{2} +\chi_1) ~~-g^*(\chi_1)}]^T, \nn \\
{\rm and } && g(\chi)~=~\cos{\chi}+i\ga_{\th_1}\sin{\chi},~~~~\ga_{\th_1}=
-ie^{i\th_1}, \nn \\
{\rm and}&& h(\chi) = \sq{1+\cos{\th_1}} ~\cos \chi + i \sq{1-\cos{\th_1}} ~
\sin \chi. \eea
This set of linear equations yields $r_A=t_N=0$,
\beq |t_A|^2 = \f{1-\cos{2\th_1}}{1+\sin^2{\th_1} +\cos^2{\th_1} ~
\sin{[2(\chi_1-\chi_2)]}}, \eeq
and $|r_N|^2 = 1 - |t_A|^2$. The cases $\vec d = \hat x$ and $\hat y$ follow
a similar calculation, and the results are given in Eqs.~\eqref{tA-p-dy} and
\eqref{rA-p-dx}.

For a $s$-wave SC, a similar calculation can be done at $E = \pm \De_0$;
once again, the existence of a bound state on the SC side with a hard wall
boundary makes two eigenspinors of the SC Hamiltonian identical. Therefore,
there are two other spinors which are orthogonal to these eigenspinors and
one can use them to reduce the problem to four equations in four unknowns
as shown above.

\end{document}